\documentclass[aps,pre,twocolumn,showpacs,floatfix]{revtex4}
\usepackage{graphicx,color}
\usepackage{amsmath,amssymb}
\usepackage{hyperref}

\newcommand{\figOne}{
\begin{figure}[b]
   \centering
   \includegraphics[width=2.25in]{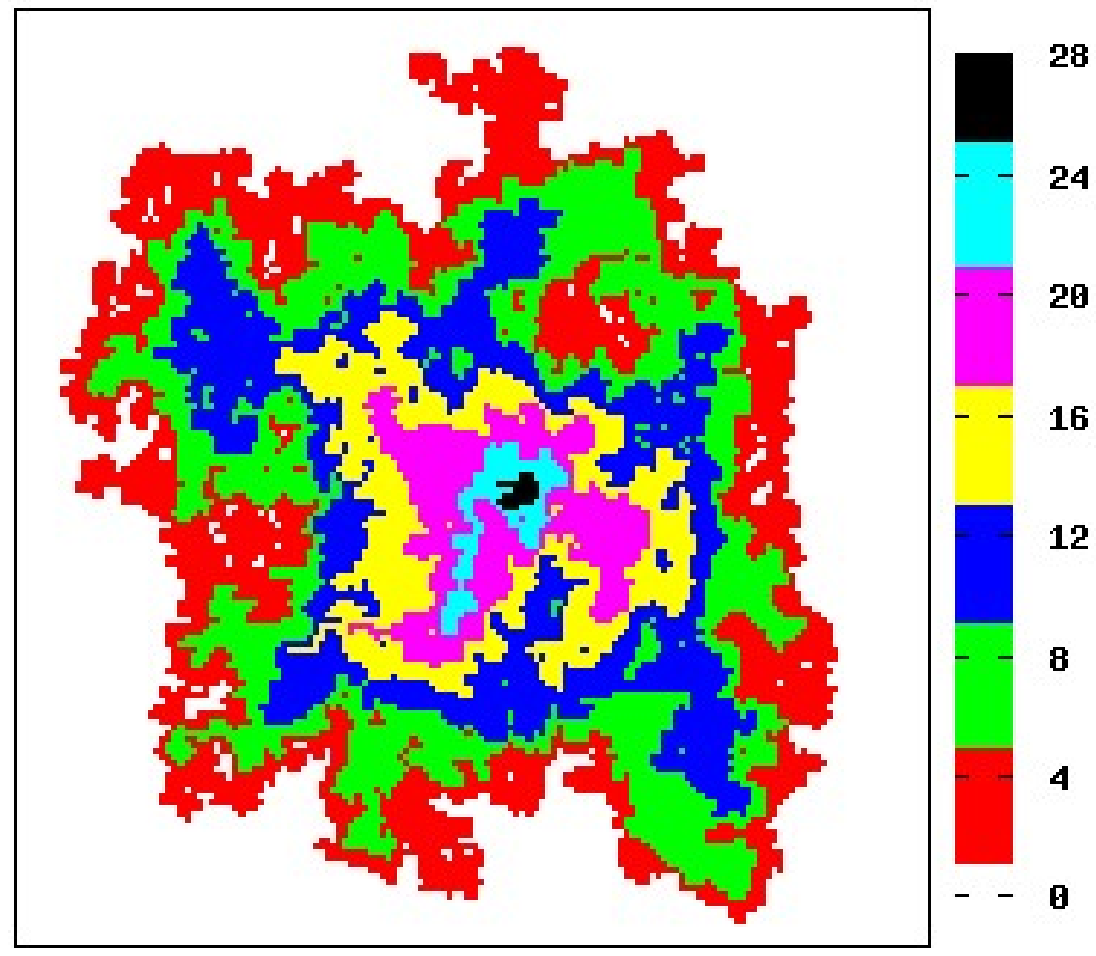}
   \includegraphics[width=2.25in]{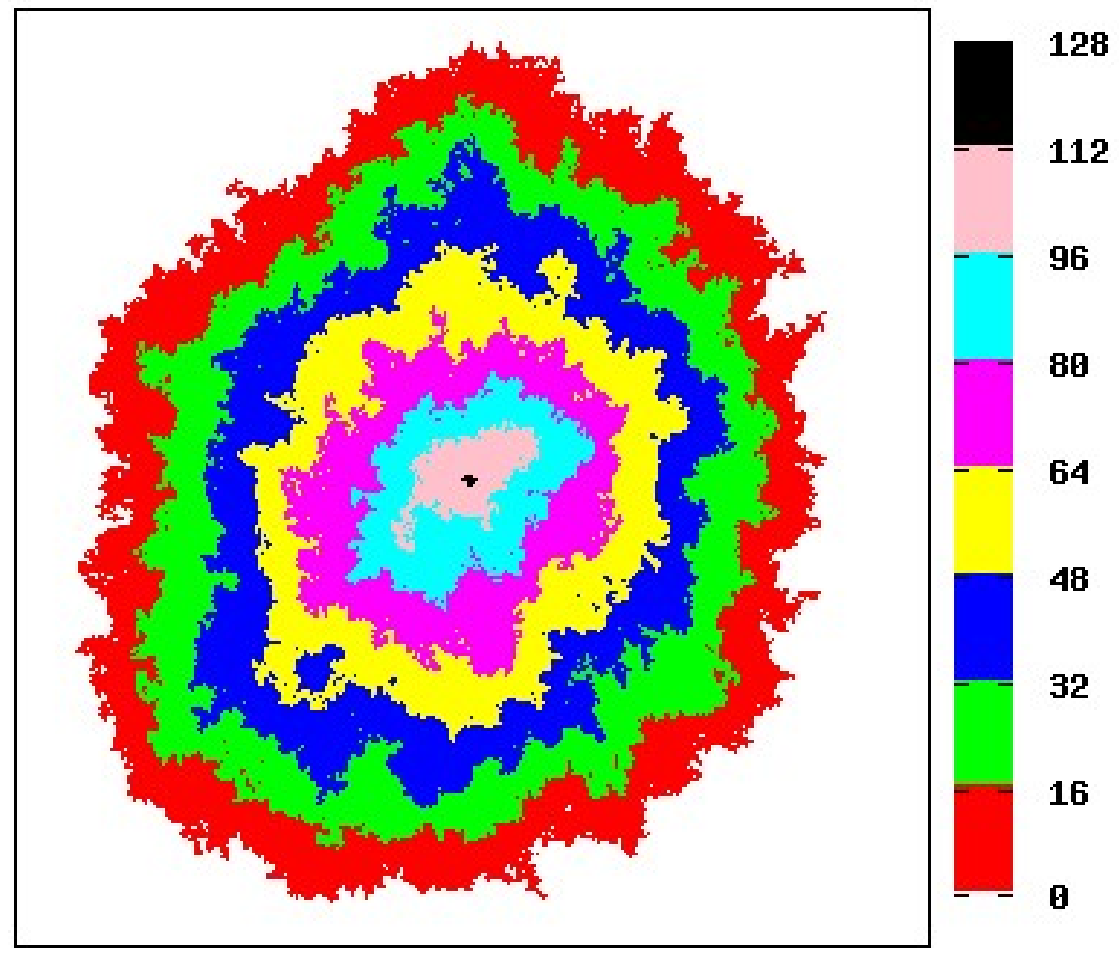}

   \caption{(Color online) Cluster formed by an EW after $10^5$ (top) and $10^7$
   (bottom) steps. A site is assigned a color depending on whether the numbers
   of visits to it is in the range $1-4, 5-8$ etc. for $10^5$, and $1-16, 17-32$
   etc. for $10^7$ steps (key panel displayed on the right). The diameters of
   the clusters are $134$ and $600$ units respectively.  }\label{fig:1}
   
\end{figure}
}

\newcommand{\figTwo}{
\begin{figure}[tb]
	\centering
	\includegraphics[width=2.5in]{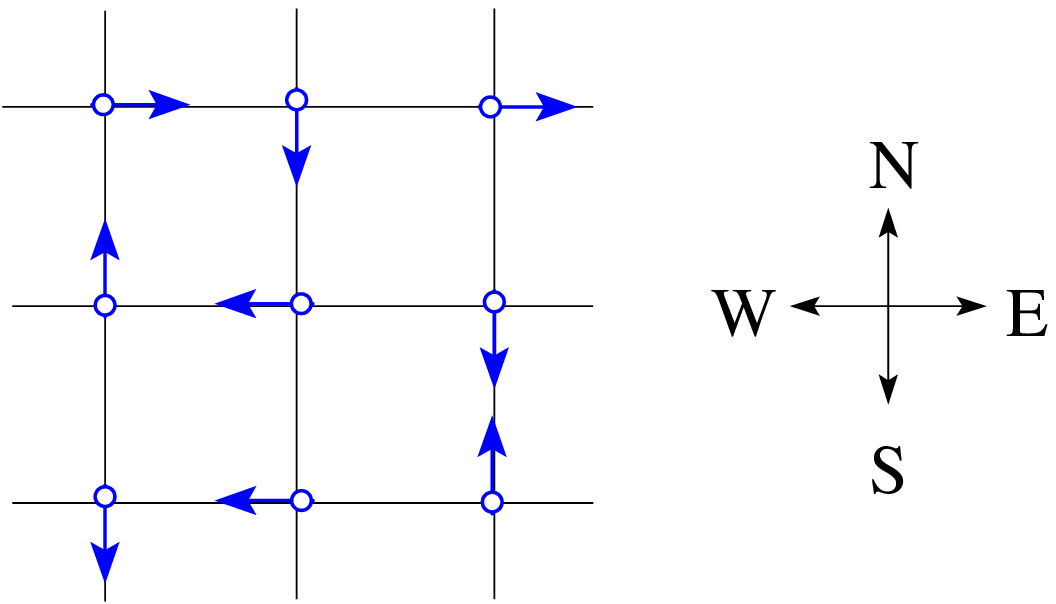}

   \caption{(Color online) A portion of the square lattice. At each lattice
   site, the outgoing bond is shown by the arrow. } \label{fig:2}

\end{figure}
}

\newcommand{\figThree}{
\begin{figure}[tb]
	\centering
	\includegraphics[width=3.45in]{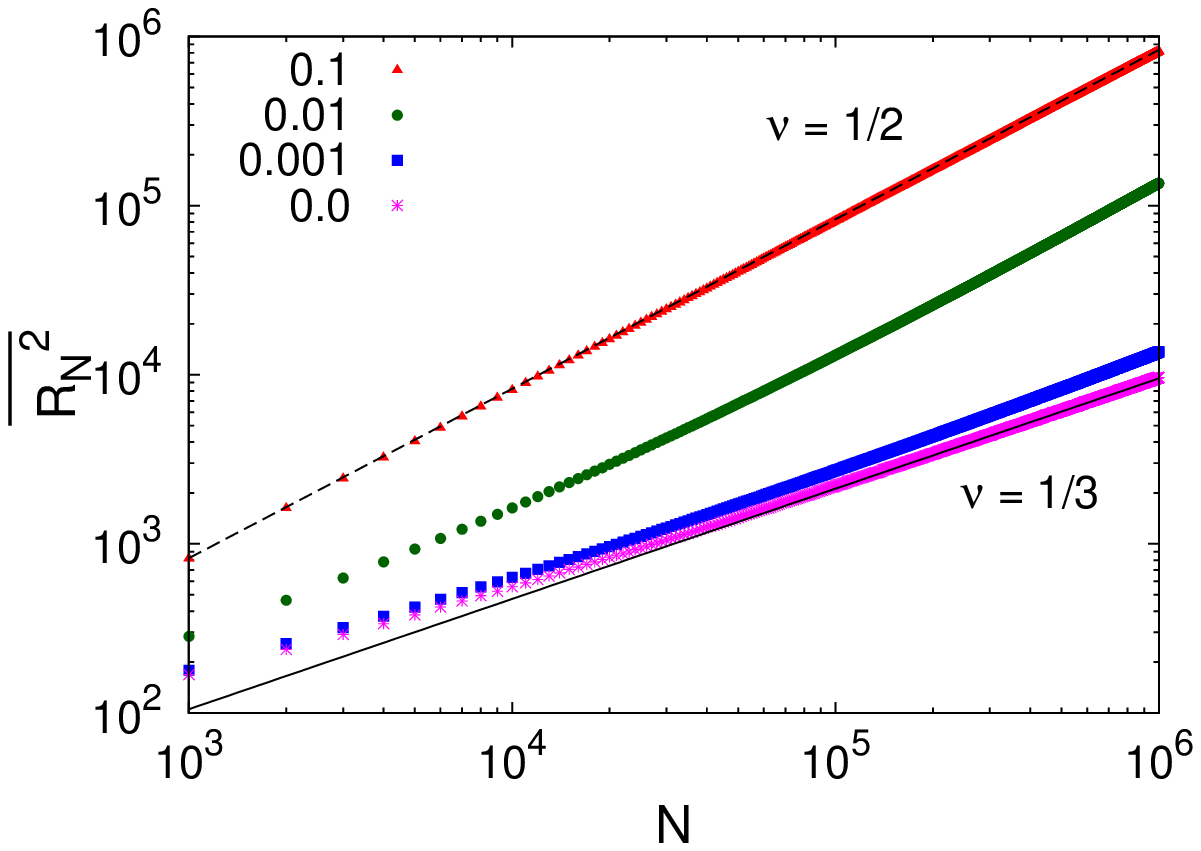}

   \caption{(Color online) Mean square displacement $\overline{R_{N}^2}$ of the
   walker from the origin as a function of $N$ for various $\epsilon$ values.
   For the pure case (i.e. $\epsilon=0$) $\overline{R_{N}^2}$ grows as
   $N^{2\nu}$ with $\nu=1/3$. In the presence of noise, there is a cross over
   from the Eulerian ($\nu =1/3$) to a simple random walk behaviour ($\nu =
   1/2$).} \label{fig:3}

\end{figure}
}

\newcommand{\figFour}{
\begin{figure}[tb]
	\centering
	\includegraphics[width=3.45in]{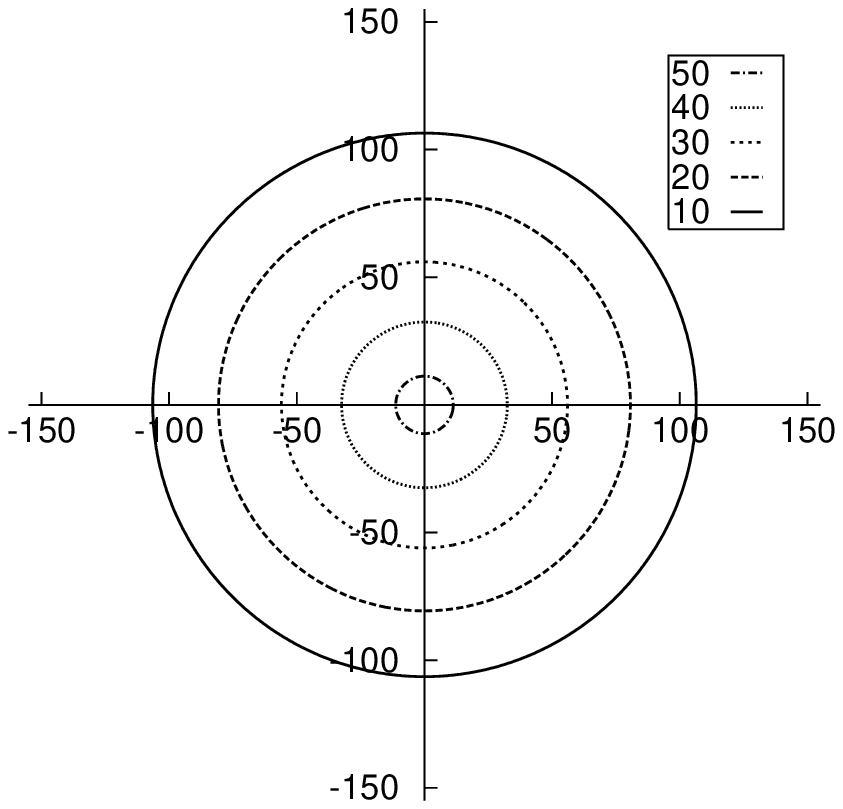}

   \caption{Contours of the constant average number of visits,
   $\overline{n_N}({\bf x})$ for various values as indicated, for the
   EW of $N = 10^6$ steps averaged over $10^7$ samples.
   }\label{fig:4}

\end{figure}
}

\newcommand{\figFive}{
\begin{figure}[tb]
	\centering
	\includegraphics[width=3.25in]{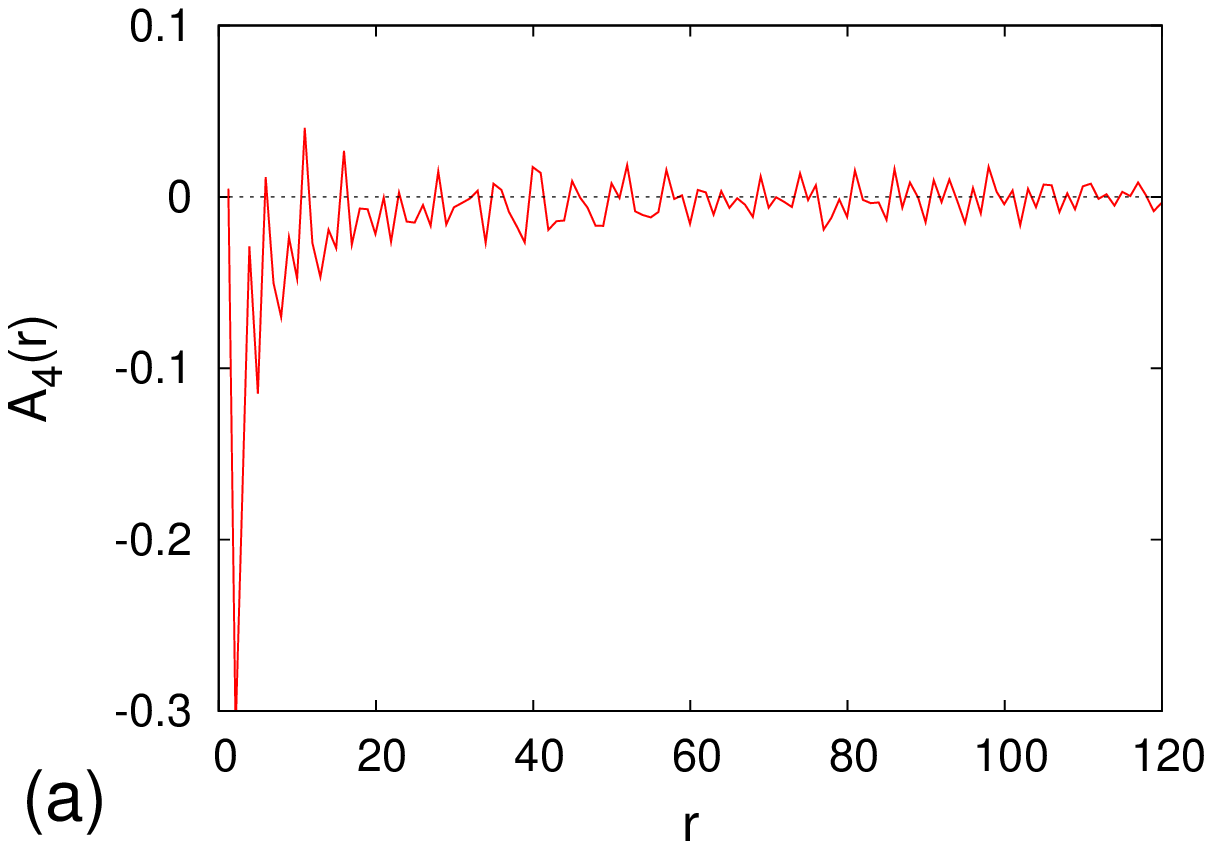}
   \includegraphics[width=2.25in]{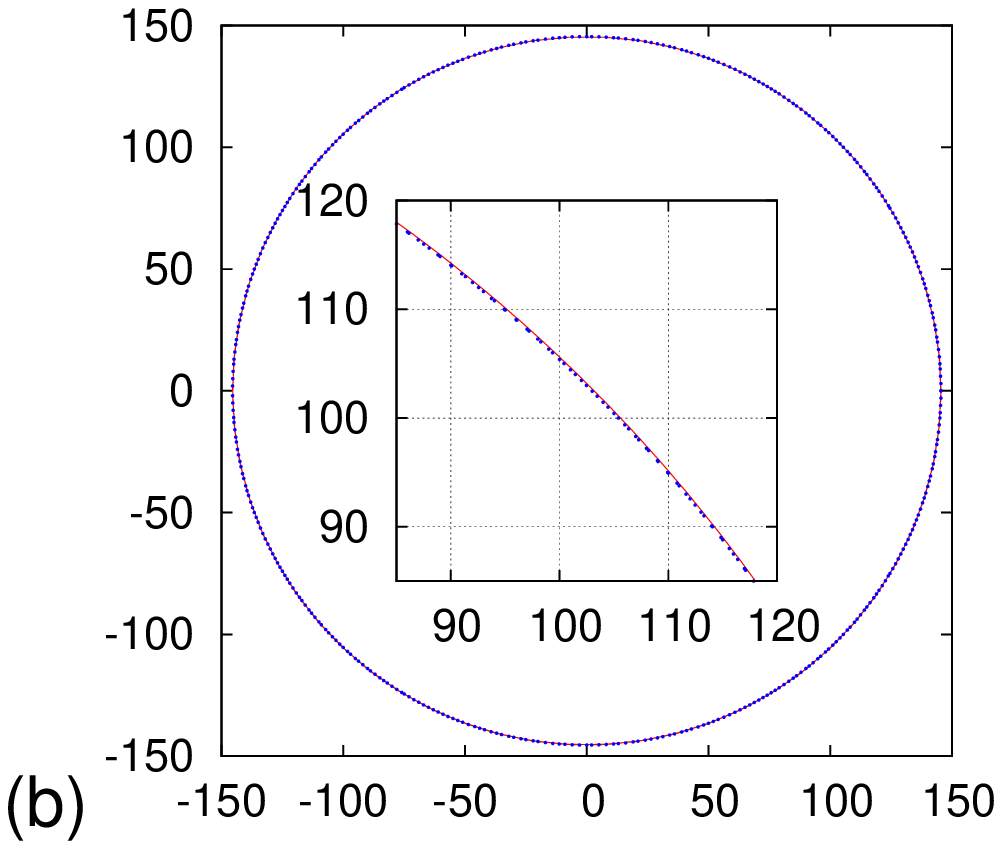}
	\includegraphics[width=3.25in]{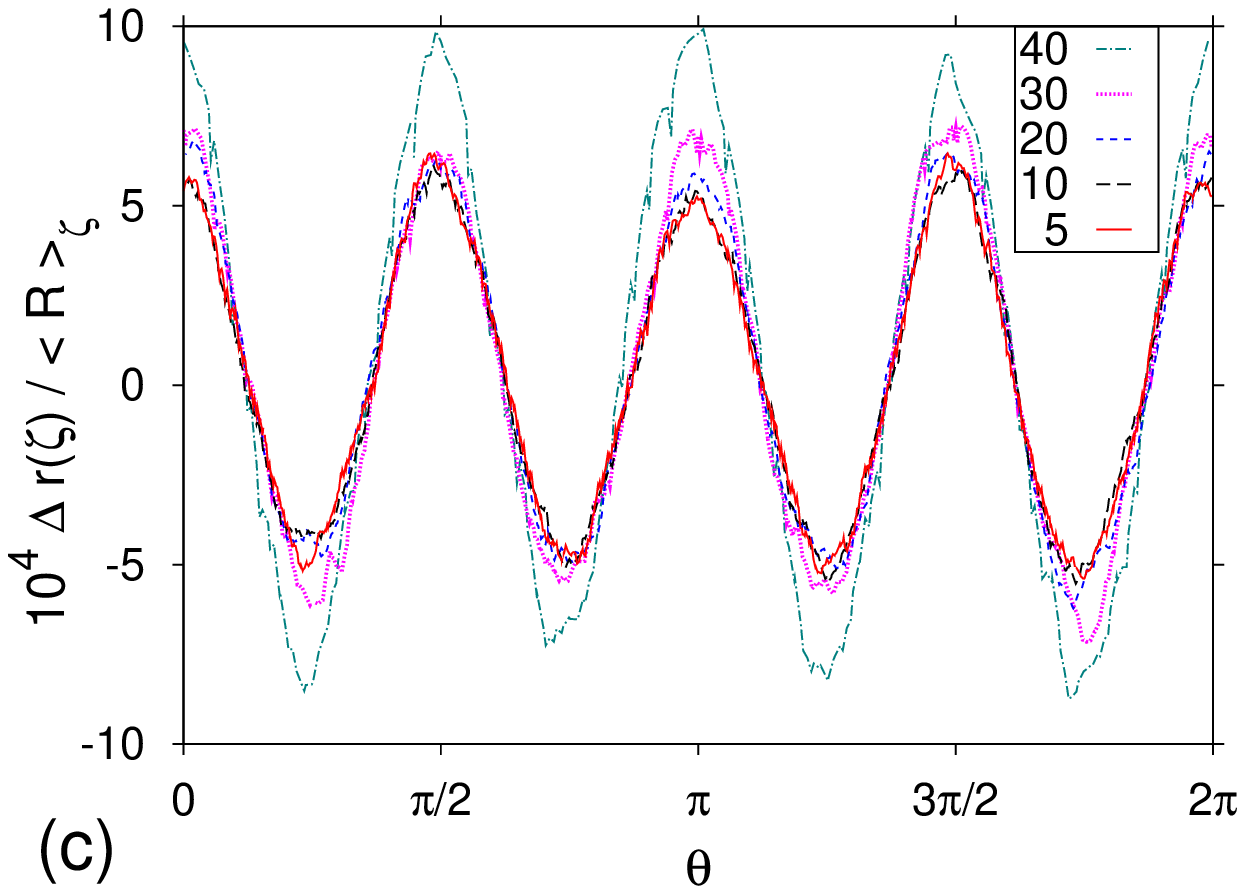}

   \caption{(Color online) (a) $A_{4}(r)$ vs $r$ for $N=10^{6}$ steps. (b)
   Contour of $\overline{n_N}({\bf x}) = 1$ and its fit. The inset zooms a
   particular portion of the curve. (c) $\Delta r(\zeta) / \langle R
   \rangle_\zeta$ as a function of $\theta$ for various $\overline{n_N}({\bf x})
   = \zeta$ values.  }\label{fig:5}

\end{figure}
}

\newcommand{\figSix}{
\begin{figure}[t]
	\centering
	\includegraphics[width=3.25in]{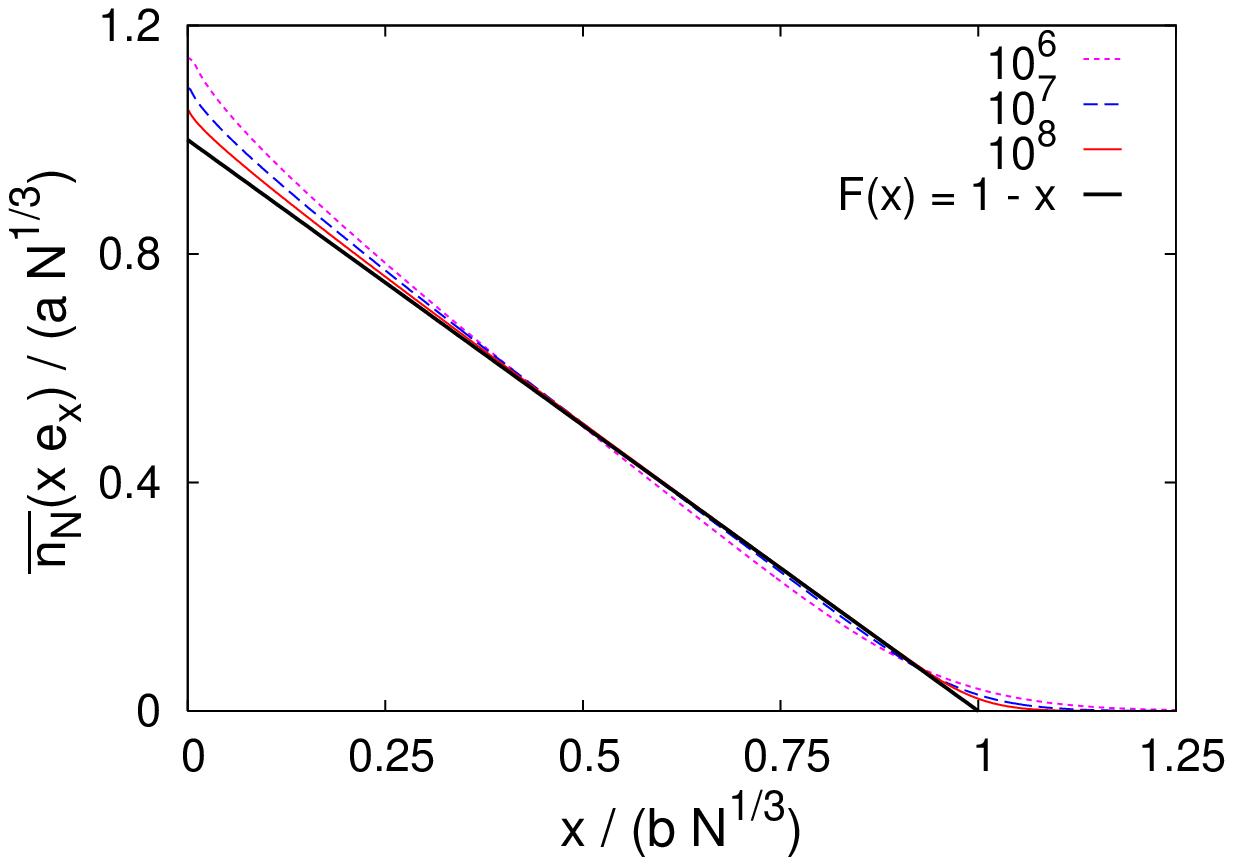}

   \caption{(Color online)Data collapse of $\overline{n_{N}}(x{\bf e}_x)$ for
   $N=10^6, 10^7$ and $10^8$ steps. The thick solid line is the scaling
   function.}\label{fig:6}

\end{figure}
}

\newcommand{\figSeven}{
\begin{figure}[b]
	\centering

	\includegraphics[width=3.25in]{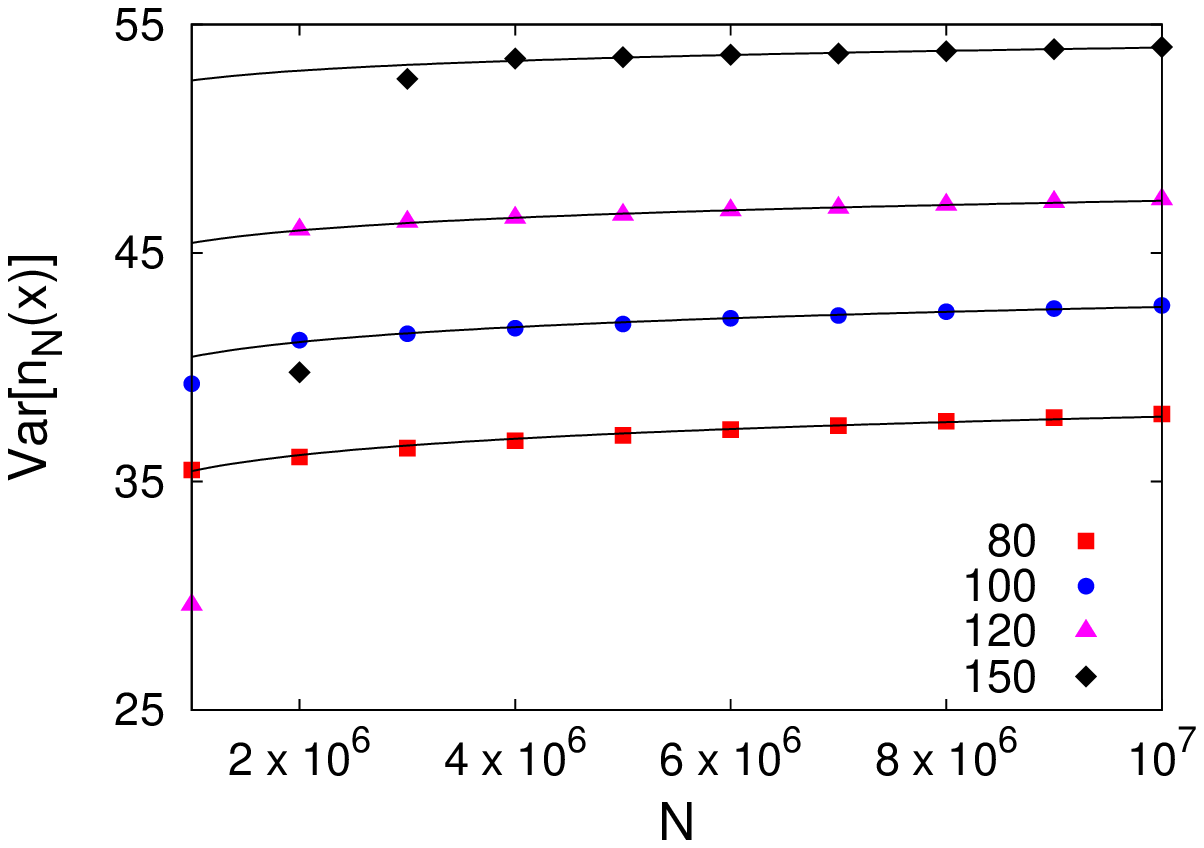}
	
   \caption{(Color online) $Var\left[ n_{N}({\bf x}) \right]$ as a function of
   length $N$ for various $|{\bf x}|$. The lines are the best fit to the data.}
   \label{fig:7}

\end{figure}
}

\newcommand{\figEight}{
\begin{figure}[bt]
	\centering
	\includegraphics[width=3.25in]{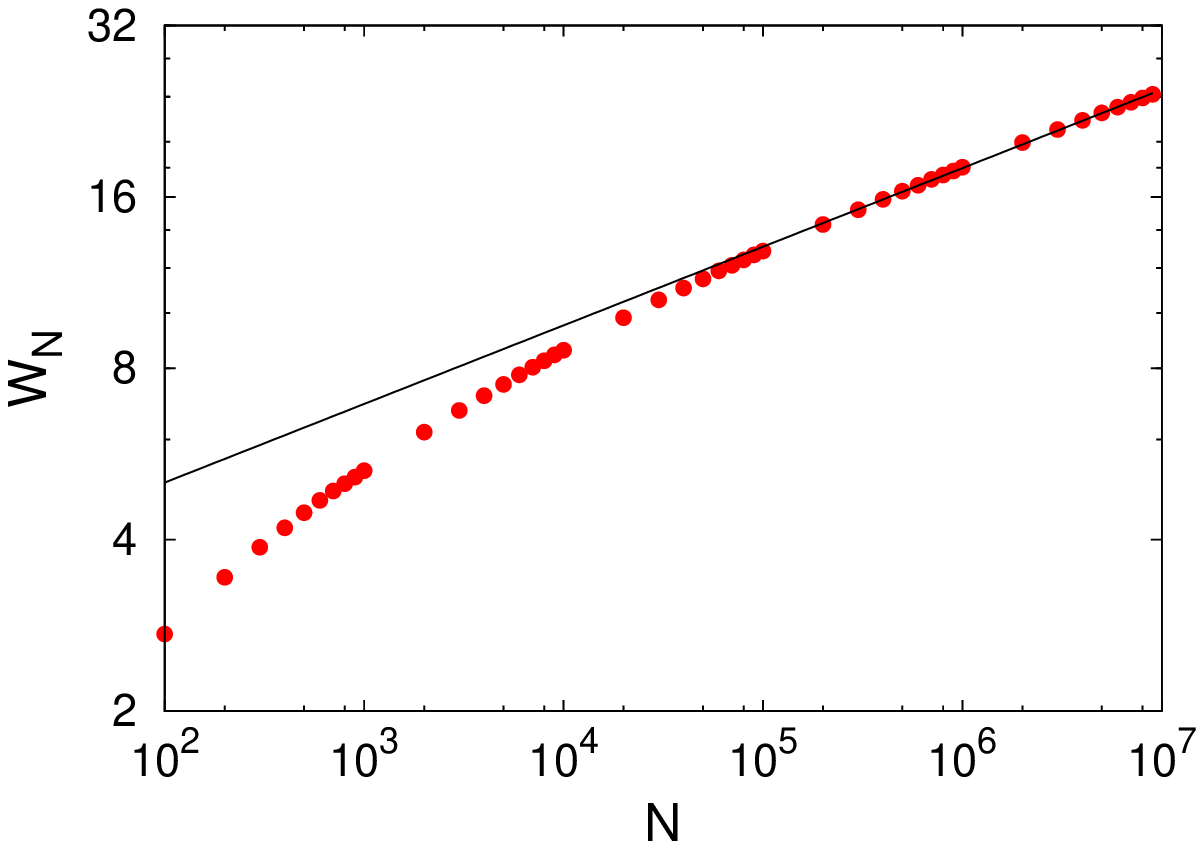}

   \caption{(Color online) The width, $W_{N}$, of the boundary of the cluster
   formed by the EW vs its length $N$. The data is averaged over $10^4$ samples.
   The straight line is the best fit to the data with slope $\delta = 0.136 \pm
   0.002$. }\label{fig:8}
	
\end{figure}
}

\newcommand{\figNine}{
\begin{figure}[t]
   \centering
   \includegraphics[width=3.25in]{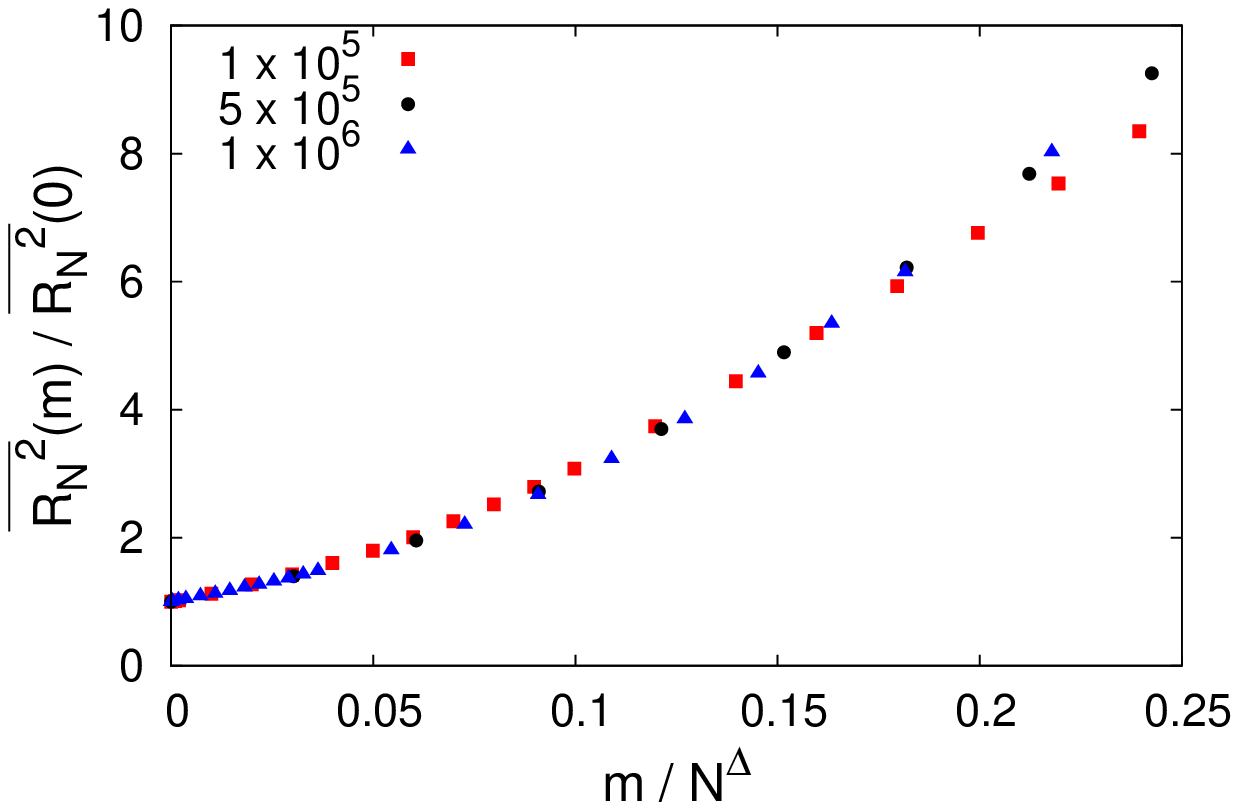}

   \caption{(Color online) $\overline{ R_N^2}(m) / \overline{ R_N^2}(0)$ vs
   $m/N^{\Delta}$ with $\Delta = 0.74 \pm 0.01$ for various $N$.} \label{fig:9}
   
\end{figure}
}

\newcommand{\figTen}{
\begin{figure}[t]
	\centering
	\includegraphics[width=2.0in]{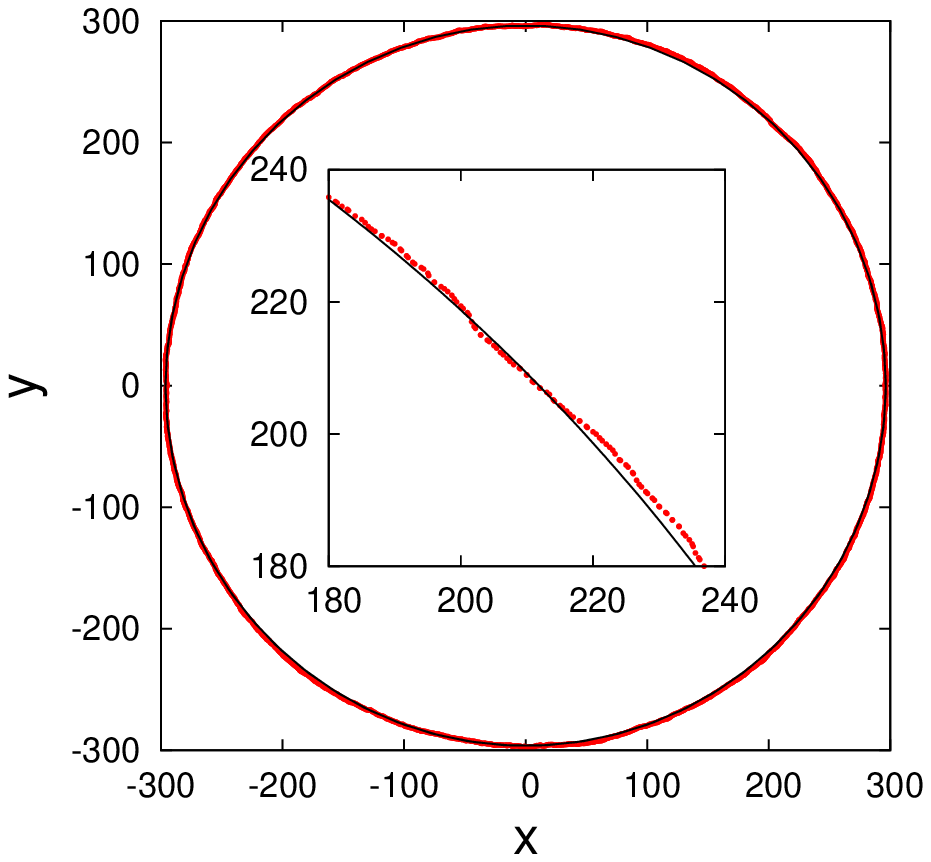}
	
   \caption{(Color online) Contour of $\overline{n_N}({\bf x}) = 1$ for the EW
   of $N=10^5$ steps in the presence of noise ($\epsilon = 0.01$). The averaging
   is done over $10^6$ samples. The line is the best fit to the data.
   }\label{fig:10}

\end{figure}
}

\newcommand{\figEleven}{
\begin{figure}[t]
   \centering
   \includegraphics[width=3.25in]{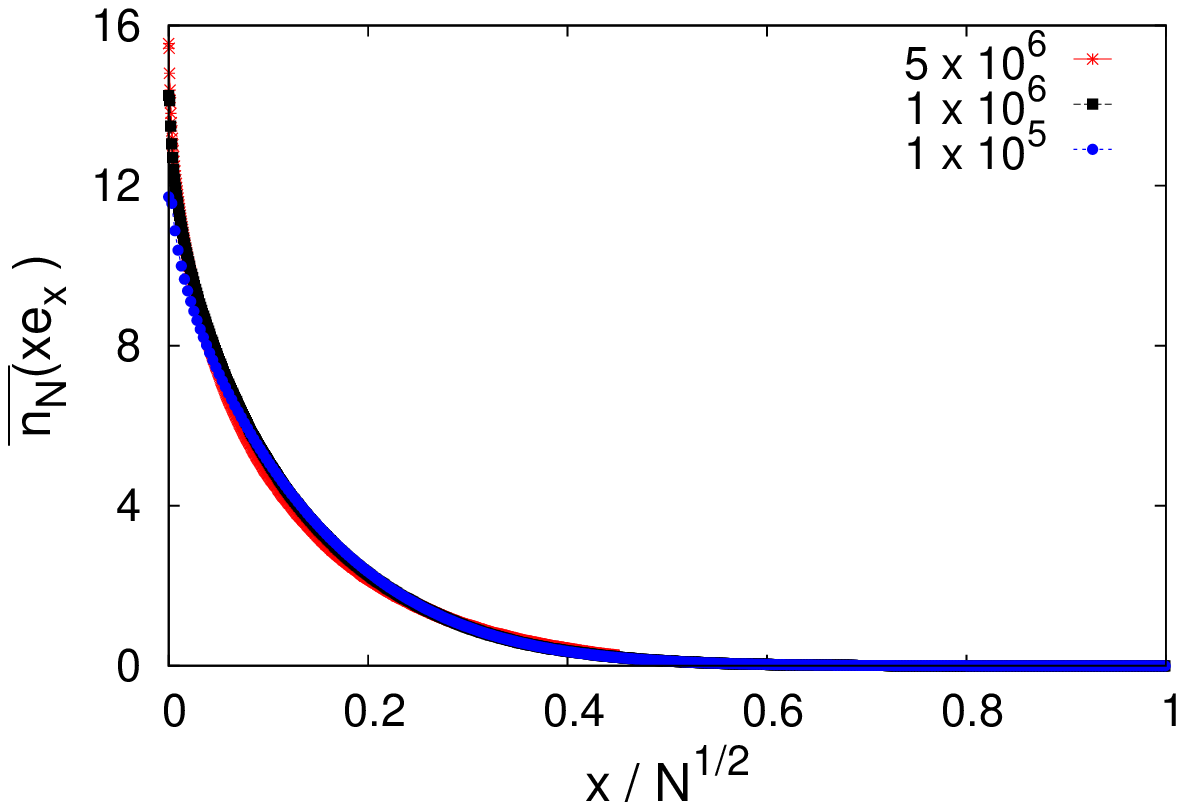}

   \caption{(Color online) Data collapse of $\overline{n_{N}}(x{\bf e}_{x})$ vs
   $x / N^{1/2}$ for the EW of $N= 1\times 10^5$, $1\times10^6$ and
   $5\times10^6$ steps with noise strength $\epsilon=0.01$.} \label{fig:11}
   
\end{figure}
}

\begin{document}
\title{Asymptotic shape of the region visited by an Eulerian Walker}

\author{Rajeev Kapri\footnote{Present address: Department of Physics,
Indian Institute of Science Education and Research Mohali, MGSIPAP
Complex, Sector 26, Chandigarh -- 160 019, India.}}
\email{rkapri@theory.tifr.res.in}
\author{Deepak Dhar}
\email{ddhar@theory.tifr.res.in}

\affiliation{Department of Theoretical Physics, Tata Institute of
Fundamental Research,\\ 
1 Homi Bhabha Road, Colaba, Mumbai -- 400 005, India.}

\date{\today}
\begin{abstract}

We study an Eulerian walker on a square lattice, starting from an initial
randomly oriented background using Monte Carlo simulations.  We present evidence
that, for large number of steps $N$, the asymptotic shape of the set of sites
visited  by the walker is a perfect circle.  The radius of the circle increases
as $N^{1/3}$, for large $N$, and the width  of the boundary region grows as
$N^{\alpha / 3}$, with $\alpha = 0.40 \pm .06$. If we introduce stochasticity in
the evolution rules, the mean square displacement of the walker, $\langle
R_{N}^{2} \rangle \sim N^{2\nu}$, shows a crossover from the Eulerian ($\nu =
1/3$) to a simple random walk ($\nu=1/2$) behaviour.

\end{abstract}


\maketitle

\section{Introduction}

There has been a lot of interest in the study of asymptotic shape of growing
clusters. One of the earliest of such studies was by Richardson, who showed that
the asymptotic shape of the infected region in an epidemic model has linear
segments ~\cite{Richardson73}. In the Eden model \cite{Eden}, which models an
epidemic without recovery, it has been shown that the  asymptotic shape of the
growing cluster is not a perfect circle~\cite{DDhar}. The shape of growing
clusters has also been studied in sandpile models. In the abelian sandpile
model, in two dimensions in an initial background of $h$ particles at each
lattice site \cite{Boer08, sadhu09}, it was found that the cluster in general
has a convex asymptotic shape, which becomes more circular as $h$ is decreased,
and tends to a perfect circle as $h\rightarrow -\infty$.

In this paper, we study the Eulerian walker (EW) model on the square lattice.
This model is related to the sandpile model and was initially introduced by
Priezzhev {\it et al.}~\cite{Priezzhev96,Priezzhev98} as a simpler variant of
the sandpile model of self-organized criticality (SOC)~\cite{BTW}. It has
subsequently found applications in design of derandomized simulations of Markov
chains~\cite{Propp09}, efficient information transfer protocols in computer
networks~\cite{Panagiotouk09}, and modelling coevolution of virus and immune
systems~\cite{Izmailian07}.  We study the model, starting from a disordered
background, by Monte Carlo simulations. We have a single walker that moves on
the lattice and we look at the shape of region visited by it which grows with
the length $N$ of the walk.  Interestingly, we find evidence that this region is
asymptotically a perfect circle. The circular shape is reminiscent of the
circular shape in the rotor-router aggregation model studied by Propp (see
\cite{Levine02,Kleber05}), where, for a special initial configuration, the
region is almost a perfect circle~\cite{Levine05} with departures from the
circle being of order 1.  The circular shape for the EW cluster is not very
evident for small walks. For example, in Fig.~\ref{fig:1}, we have shown
clusters formed by the EW of $N=10^5$ and $10^7$ steps. Clearly, only for large
$N$, does the circular shape start to emerge, and it requires careful
statistical analysis to see this when $N$ is not so large.

\figOne

The EW model can also be looked upon as a particular limit of a growing
self repelling walk, in which the walker preferably jumps along a bond
which has been visited least number of times so far.  This model  was
studied in $1d$ by Toth and Veto~\cite{Toth08}. In the zero temperature
limit, in one specific variant, this becomes the EW model and a finite
temperature corresponds to noise.  It was found that the number of
visits at a distance $y$ from the origin satisfies the scaling function
$F(y) = 1 - y, \ \text{for} \ 0 \le y \le 1$. We find that the same scaling
function holds even in $2d$. We also study the model in the presence of
noise, where there is a small probability $\epsilon$ that walker goes in
a direction not given by the EW rule. We find that a small noise changes
the asymptotic behaviour. The diameter of growing region scales as
$A(\epsilon)N^{1/2}$ in the presence of noise and as $N^{1/3}$ in its
absence.

The paper is organized as follows. In Sec.~\ref{sec:model}, we define
our model. In Sec.~\ref{sec:case1}, we give details of the simulation
without noise (i.e. $\epsilon=0$). We find that the asymptotic shape of
lines of constant average number of visits by an EW are perfectly
circular, within statistical errors. We will argue that the variance of
average number of visit at any distance from the origin tends to a
finite number for large $N$ and also obtain the scaling function for the
average number of visits. In Sec.~\ref{sec:case2} we discuss the  case
with noise, and finally we summarize our results in Sec.~\ref{sec:conc}. 

\section{Model} \label{sec:model} 

\figTwo

The Eulerian walker is defined as follows: We consider a square lattice.
We associate with each site an arrow which can point to along one of the
four directions, denoted by N, E, S and W (Fig.~\ref{fig:2}). In the
initial configuration, the direction of the arrow at each site is chosen
independently, and with equal probability. We put a walker at the origin
which moves on the lattice. The motion of the walker is affected by
configuration of arrows on the lattice, which in turn affects the arrow
configuration on the lattice.  The walker follows the following rule: at
each time step, the walker after arriving at a site rotates the arrow at
that site in a clockwise direction by $90^{\circ}$, and then moves one
step along the new arrow direction.

It was shown in Ref. \cite{Priezzhev96} that on any finite graph, using the
above rules, the walker eventually visits all sites and settles into a limit
cycle which is an Eulerian circuit visiting each directed bond exactly once in a
cycle.  This is not the case on an infinite lattice, where the walker always
finds new bonds which are not visited earlier and the number of visited sites
keeps on growing. It was noted already that in $2d$, the diameter of the region
visited by the walker grows as $N^{1/3}$, but the asymptotic shape was not
investigated. 

The EW with noise is defined as follows: at each time step the walker rotates
the arrow at its location by $0^{\circ}$, $90^{\circ}$, $180^{\circ}$ or
$270^{\circ}$ with probability $\epsilon/3$, $1-\epsilon$, $\epsilon/3$ or
$\epsilon/3$ respectively.  We will show that the diameter of the region visited
grows as $N^{1/2}$ for nonzero $\epsilon$ and as $N^{1/3}$ when $\epsilon=0$.

\section{Numerical simulations for evolution without noise}
\label{sec:case1}

First we discuss the case $\epsilon=0$. We denote the number of times different
sites visited by the walker after $N$ steps by $n_{N}({\bf x})$ and the walker's
square displacement from the origin by $R_{N}^2$. We evaluate
$\overline{n_{N}}({\bf x})$ (the over line represents averaging over initial
conditions), the variance of $n({\bf x})$ denoted by $Var\left[n_{N}({\bf
x})\right]$, and the mean square displacement $\overline{R_N^2}$ by averaging
over $10^6$ different initial configurations. 

\figThree

\subsection{Mean square displacement}

According to the heuristic argument given in Ref. \cite{Priezzhev96}, if at
time $t$ the number of sites visited by the walker is $S(t)$, then in
the previous $4S(t)$ time steps, most of these sites have been visited
exactly $4$ times except a small fraction at the boundary. As the
cluster is seen to have few holes, it is nearly compact, and   $S(t)
\sim D^{2}(t)$, where $D(t)$ is the diameter of the cluster, at time $t$.
Thus we get
\begin{equation}
	\frac{dD(t)}{dt} \sim \frac{1}{D^2},
\end{equation}
which implies that after $N$ steps,
\begin{equation}
	\label{eq:rrms}
   D_{N} \sim N^{\nu} \quad \text{with}\ \nu = \frac{1}{3}.
\end{equation}

Figure~\ref{fig:3} shows the mean square displacement of the EW as a
function of its length $N$. The averaging is done over $10^6$
realizations. The straight line, which is the best fit to the data has a
slope $0.33 \pm 0.01$, consistent with Eq.~(\ref{eq:rrms}).

\subsection{Average number of visits}

As seen in Fig.~\ref{fig:1}, the cluster of visited sites is quite
irregular in shape. Also sites that have been visited at least $n$ times
have rough boundaries with several islands of sites that have been
visited fewer number of times than all the surrounding sites. However,
if we average over different realizations of the initial arrow
configuration, some interesting regularities are seen.

In Fig.~\ref{fig:4}, we have plotted lines of $\overline{n_{N}}({\bf x})
= \zeta$, for different $\zeta$ as indicated in the figure, for an EW of
length $N = 10^6$ averaged over $10^7$ realizations.  To obtain these
lines, we add a diagonal bond between $(x,y)$ and $(x+1,y+1)$ for each
$(x,y)$, and extend the definition of $\overline{n_N}({\bf x})$ to all
real ${\bf x}$ by linear interpolation within each small triangle. The
plot shows that these lines are nearly perfect circular in shape. 

\figFour

The shape of rings, for large $N$, can be defined by a function
\begin{equation}
   f(\theta) = \lim_{N \rightarrow \infty} \frac{r_N(\theta)}{ N^{1/3}},
\end{equation}
where $r_{N}(\theta) \ (0 \le \theta < 2\pi)$ is the angle dependent
radius. If the shape is a perfect circle $f(\theta) = constant$,
otherwise $f(\theta)$ is a periodic function of $\theta$ that can be
expressed in terms of Fourier cosine series
\begin{equation}
	\label{eq:fseries}
	f(\theta) = \sum_{m = 0}^{\infty} a_{4 m} \cos( 4 m \theta).
\end{equation}
Since the shape has fourfold symmetry the series will only have terms
with $m=4u \ (u=0,1,2\dots)$. The vanishing of $a_{ 4m}$'s  for all $m
\neq 0$ then implies a circular shape. We define
\begin{equation}
	A_{4}(r) = \frac{\sum_j \overline{n_N}({\bf x}_j)
	\cos(4\theta_j) }{ \sum_j \overline{n_N}({\bf x}_j)}
\end{equation}
as the normalized amplitude of the fourth Fourier mode. The summation
$j$ is over all the lattice points whose Euclidean distance from the
origin lies between $r$ and $r+1$. Here $\theta_j$ is the angle that the
vector ${\bf x}_j$ makes with the $x$-axis. This function, for a fixed
$r$, has a well defined limit for $N \rightarrow \infty$. In
Fig.~\ref{fig:5}(a), we have shown $A_4(r)$ as a function of $r$ for
$N=10^6$ steps. The plot shows that for large $r$, $A_{4}(r)$ approaches
zero with fairly large irregular-looking fluctuations. These
fluctuations, for a fixed $r$, \emph{do not become smaller by statistical
averaging or larger $N$}. These fluctuations occur because the lattice points
lying between radii $r$ and $r+1$ are not distributed perfectly evenly along the
ring. They are of number-theoretic origin, and have been
studied in the mathematics literature under the name of the `Gauss
circle problem'~\cite{Grosswald84}. The analysis of moments of $A_4$ is
therefore not very useful to estimate the shape and we have to adopt
some other procedure.

\figFive

The simplest way to have a quantitative estimate of shape of rings is to
fit the data and obtain the mean radius for various rings of constant
$\overline{n_N}({\bf x})$. In Fig.~\ref{fig:5}(b), we have shown such a
ring and its fitting for $\overline{n_N}({\bf x}) = 1$. The best fit
gives the mean radius $\langle R \rangle = 145.436 \pm 0.003$ for
$N=10^6$. For other rings also we find the error bars of the same order
showing that the line of constant average number of visits are circular
in shape within an error bar of $0.002\%$. The inset shows a close up of
a particular region of the ring. 

We also calculate the root mean square deviation of distance, $\Delta
r(\zeta)$, of points on the line of constant $\overline{n_N}(x)=\zeta$
to the origin with mean radius $\langle R \rangle_{\zeta}$. This is
shown in Fig.~\ref{fig:5}(c) as a function of $\theta$ for various
$\zeta$. The plot shows that, $\Delta r(\zeta)/\langle R
\rangle_{\zeta}$ is of the order $10^{-4}$ and decreases as $\langle R
\rangle_{\zeta}$ increases. 

Another way to estimate the shape of the cluster formed by visited sites
is to obtain various moments of the data. Since all the four directions
are equivalent for the walker, we expect that $\overline{n_N}({\bf x})$
has a fourfold symmetry. For a given length $N$, we calculate $\langle
x^4 \rangle$, $\langle y^4 \rangle$ and $\langle x^2 y^2 \rangle$
moments. If the shape of the cluster is perfect circular we would have
\begin{equation}
   \label{Eq:8}
	\frac{ \langle x^4 \rangle}{ \langle x^2 y^2 \rangle} =
	\frac{\langle y^4 \rangle}{ \langle x^2 y^2 \rangle} = 3.
\end{equation}
For $N=10^6$ steps averaged over $10^6$ initial realizations, we find
that $\langle x^4 \rangle / \langle x^2 y^2 \rangle = \langle y^4
\rangle / \langle x^2 y^2 \rangle = 3.007$, which is consistent with the
asymptotic value $3$ deviation being only about $0.2\%$.

\subsection{Scaling of $\overline{ n_{N}}({\bf x})$}

For large $N$, $\overline{ n_{N}}({\bf x})$, the average number of
visits to the site ${\bf x}$,  depends only on $|{\bf x}|$. Therefore,
we expect that $\overline{ n_{N}}({\bf x}) $, satisfies the scaling form
\begin{equation}
	\label{eq:FSS}
   \overline{ n_{N}}(|{\bf x}|) = a N^{1/3} F \left( \frac{ |{\bf x}|}
   {b N^{1/3}} \right),
\end{equation}
where $F(y)$ is the scaling function.

The scaling function can be determined as follows: Let
${\bf x}_1$ and ${\bf x}_2 \ (|{\bf x}_2| > |{\bf x}_1|)$ be the
distances of two different sites from the origin. The walker would have
made several visits to ${\bf x}_1$ before it first reaches ${\bf x}_2$.
Afterwards,  because of the local Euler-like organization of the arrows,
both sites are visited equally often. Therefore, the difference between
the number of times sites at distances ${\bf x_1}$ and ${\bf x_2}$ are visited
remains bounded as $N\rightarrow \infty$, i.e.,
\begin{equation}
   \overline{n_N}({\bf x}_1) - \overline{n_N}({\bf x}_2) = a N^{1/3}
   \left[ F(y_1) - F(y_2) \right] = constant.
\end{equation}
This implies that $F(y)$ must be a linear function of $y$. Using the
freedom of choice of constants $a$, and $b$, we can set $F(0) = 1$
and $F(1) = 0$. Therefore, we have 
\begin{equation}
	\label{eq:fx}
	F(y) = \begin{cases} 1 -  y \quad {\rm for} \ 0 \le y \le 1, \cr
		0 \quad {\rm otherwise}.
	\end{cases}
\end{equation}
This simple form of $F(y)$  was already noted by Toth and Veto for
the problem in one dimension~\cite{Toth95,Toth08}. The normalization
condition  $\int \overline{n_N}({\bf x}) d{\bf x} = N$ gives $ab^2 = 3/\pi$.

\figSix 

In Fig.~\ref{fig:6}, we have plotted the finite size scaling of
$\overline{n_N}(x{\bf e}_x) $ for $N=10^6, 10^7$ and $10^8$ steps with
$a=0.5$ and $b=1.38$. In the same plot, we have also shown the scaling
form given by Eq.~(\ref{eq:fx}). The plot shows that as $N$ is
increased, the scaled data approaches rather slowly towards the scaling
form. In particular for smaller $|{\bf x}|$, the approach to the
asymptotic curve seems to be  slow.

\subsection{Variance of $n_{N}({\bf x})$}

\figSeven

We also monitored the variance of $n_{N}({\bf x})$ as a function of $N$
for different $|{\bf x}|$. This is plotted in Fig.~\ref{fig:7}.  The
graph shows that $Var\left[n_{N}({\bf x})\right]$ increases slowly with
$N$ and suggests that $Var[n_N({\bf x})]$ remains finite for all fixed
${\bf x}$. This can be understood as follows: the variance of
$n_{N}({\bf x})$ arises only from the randomness in the initial visits
of walker to ${\bf x}$. Once the local bonds have been organized into a
near Euler circuit, the subsequent increments in $n_N({\bf x})$ are
nearly deterministic.

\subsection{Roughness exponent}

We define the surface of the set of visited sites as all visited sites that have
at least one unvisited neighbor. Let $W_N ^2$ denotes the variance of the
distance from the origin of randomly picked surface formed by the EW of length
$N$. Then, $W_N^2$ is defined by
\begin{equation}
	\label{eq:width}
	 W_N ^2  = \overline{ \frac{1}{M} \sum_{i=1}^{M} \left(
    |{\bf x}_i| - \langle R \rangle \right)^2 },
\end{equation}
where $M$ is the number of surface points $|{\bf x}_i|$  of the cluster,
$\langle R \rangle$ is the average distance of a perimeter site, and the overbar
denotes averaging over different clusters. We define the width of the surface by
square root of $W_N^2$. 

In Fig.~\ref{fig:8}, we have plotted $W_N$ for various $N$, in a log-log scale,
averaged over $10^4$ different clusters. We observe that $W_N \sim N^{\delta}
\sim R^{3 \delta} \sim R^{\alpha}$ with $\delta = 0.136 \pm 0.02$, which gives
the roughness exponent $\alpha =  0.40 \pm 0.06$. The effective value of
$\alpha$ seems to decrease with $N$, and it is difficult to estimate its
limiting value.  It is consistent with the asymptotic value $1/3$ expected for
the Kardar-Parisi-Zhang (KPZ) surface growth process~\cite{HHZ95}. Note that in
order to think of growth of cluster of visited sites as a local growth process,
we have to redefine time so that the radius of the cluster grows linearly in the
new variable.

\figEight

\section{Numerical simulations for evolution with noise}\label{sec:case2}

For the evolution with noise also, we monitored $\overline{ R_{N}^2 }$
and $\overline{n_{N}}({\bf x})$ for various $N$ by averaging over $10^6$
different initial configurations. 

\subsection{Mean square displacement}

For zero noise, for $N < 10^3$, $\overline{ R_{N}^2 }$ increases roughly
linearly with $N$ and then as $N^{2/3}$. In Fig.~\ref{fig:3}, we have also shown
$\overline{ R_{N}^{2} }$ as a function of $N$ for noise strengths $\epsilon =
0.001$, $0.01$, and $0.1$. For $\epsilon=0.001$, there is only a small change
from the Eulerian like behaviour. However, when the noise strength is increased,
there is a clear crossover seen in $\overline{ R_{N}^{2} }$ from the Eulerian
like to a simple random walk behaviour, i.e. $ \overline{ R_{N}^{2} } \sim N$.
This crossover can be observed even for noise strength as small as
($\epsilon=0.01$).  That presence of noise changes the critical behaviour is not
so unexpected. In equilibrium critical phenomena, the well known Harris
criterion~\cite{Harris74} characterizes a large class of systems where the
critical behavior is substantially altered by the presence of disorder.  In SOC
models, the Manna model~\cite{Manna91} with stochastic toppling rules is in a
different universality class than the model with deterministic toppling rules
(e.g., the Bak-Tang-Wiesenfeld model)~\cite{BenHur96}. In fact, different types
of stochasticity can yield different universality classes. For example, in the
directed sandpile with stickiness one gets a different behaviour than the
stochastic Manna model (for the undirected case the situation is less clear
\cite{Mohanty,Munoz}). It is known that in some sandpile models, adding noise
can change the transition from first order to continuous \cite{Lee-Lee}.

\figNine

Let $m= N \epsilon$ be the number of `mistakes' the walker makes in a
walk of $N$ steps and let $\overline{ R^2_N} (m) $ represents the mean
square displacement of a walk  with $m$ mistakes. We would like to  know
how $\overline{ R^2_N} (m) $ increases with $m$. Consider the effect of
a single mistake. The rules of the Eulerian walk are such that as the
walk evolves, the initial random arrow-directions are rearranged into
an Euler circuit gradually. A wrong step would disrupt an  evolving
local Eulerian circuit. This local defect can be repaired, and a
possibly different locally Euler-like circuit formed, when the walker
revisits the site. Therefore, there is not much change in $\overline{
R^2_N} (m)$ for small $m$. However, if $m$ is  larger (i.e., of the order
$N$), the walker keeps on making new mistakes before the old mistakes
can be corrected, and the self-organization  into Eulerian circuits is
lost. This suggests that there should be a value $m^{*}$, such that
walks are nearly Eulerian for $m < m^{*}$, and random walk-like for $m >
m^{*}$.  We find that $m^{*} \sim N^{\Delta}$ with $\Delta < 1$.
Furthermore, we find that $\overline{ R^2_N} (m)$ satisfies a scaling
relation
\begin{equation}
   \label{eq:R2m}
   \overline{ R^2_N} (m)  = \overline{ R^2_N} (0) G \left(
   \frac{m}{N^{\Delta}}  \right),
\end{equation}
where $G(x)$ is the scaling function. In Fig.~\ref{fig:9}, we have
plotted  $\overline{ R^2_{N}}(m) / \overline{ R^2_N} (0)$ vs
$m/N^{\Delta}$ for $N = 1 \times 10^5$, $5 \times 10^5$ and $1 \times
10^6$ steps. A good collapse is obtained for $\Delta = 0.74 \pm 0.01$.
Equation~(\ref{eq:R2m}) can also be written as (using $\epsilon N = m$)
$\overline{ R^2_N}(\epsilon) = \overline{ R^2_N}(0) G \left( \epsilon
N^{1-\Delta} \right) $. Then we have $\overline{ R^2_N}(0) \sim N^{2/3}$
and $\overline{ R^2_N}(\epsilon) \sim N$, which implies that $G(x)$
should increase as $x^{1/3(1-\Delta)}$ and the cross over value
$N^*(\epsilon) \sim\epsilon^{-1/3(1-\Delta)}$. A direct reliable
estimate of $N^*(\epsilon)$ is not possible from our data and this
scaling theory prediction is difficult to check from our simulations.

\subsection{Average number of visits}

\figTen

In Fig.~\ref{fig:10}, we have shown the contour of $\overline{n_N}({\bf
x}) = 1$ for an EW of length $N=10^6$ steps in the presence of noise
with strength $\epsilon = 0.01$. The averaging is done over $10^6$
initial conditions. The best fit to the data gives the mean radius
$\langle R \rangle = 296.401 \pm 0.015$, i.e. circular within an error
bar of $0.005\%$. The inset shows the close up of a particular region of
the ring. On comparing this with the $\epsilon = 0$ case, we see that
the mean radius of an EW with $1 \%$ noise strength is about two times
greater than the mean radius without noise.

In Fig.~\ref{fig:11}, we have plotted $\overline{n_N}(x {\bf e}_x)$ vs
$x/N^{1/2}$, where ${\bf e}_x$ is the unit vector along $x$-axis, for an
EW in the presence of noise of strength $\epsilon = 0.01$ for $N
=1\times 10^5$, $1\times 10^6$ and $5\times 10^6$ steps. The plot shows
that, except near the origin where there is a $\log(N)$ dependence, the
data for various $N$ collapse on top of each other.

\figEleven

We also obtain the width, $W_N$, of the surface formed by the EW in the presence
of noise. We find that $W_N \sim N^{1/2} \sim R$ as expected for random walks.
Hence for nonzero $\epsilon$, the asymptotic shape of the cluster is not
circular in individual realizations. Some circular symmetry is only seen in the
ensemble averages (i.e., lines of constant $\overline{n}({\bf x})$).

\section{Conclusions}\label{sec:conc}

We have studied an Eulerian walker on a $2$ dimensional square lattice
using Monte Carlo simulations. In the absence of noise, the mean square
displacement $\langle R_N^2 \rangle \sim N^{2/3}$. We find that lines of
constant average number of visits seem to be  perfectly circular. This
result is not entirely unexpected, given the fact that Eulerian walkers
can be considered as a particular type of derandomized random walkers
\cite{Propp09}. However, note that  the result does depend on the fact
that Eulerian walkers in two dimension return to origin infinitely
often, and while the {\it averaged} cluster shape would be expected to
show circular symmetry, one does not expect {\it each} individual
cluster to show spherical shape in dimensions $d>2$.

We also estimated the roughness exponent for the  boundary of the
visited region. This has a slow convergence to its asymptotic value, but
our data is consistent with it belonging to the KPZ universality class.
We also find that  even a small randomness in the  rule for walker's
next step changes the asymptotic properties:  the mean square
displacement shows a crossover from the Eulerian like  $\langle R_N^2
\rangle \sim N^{2/3}$ to a simple random walk behaviour $\langle R_N^2
\rangle \sim N$.  In higher dimensions $d > 2$, an Eulerian walker does
not return to previously visited sites often, and one would expect a
random walk like behavior even in the zero-noise case.

DD thanks B. Toth and B. Veto for discussions. We thank J. Propp for a critical
reading of the earlier version of the paper.

\end{document}